\documentclass{ws-procs9x6-cpt22}
\begin{document}

\newcommand{\refeq}[1]{(\ref{#1})}
\def\etal {{\it et al.}}

\title{New Developments in the Hamiltonian Formulation\\ of the Gravitational Standard-Model Extension}

\author{M.\ Schreck}

\address{Departamento de F\'{i}sica, Universidade Federal do Maranh\~{a}o,\\ Campus Universit\'{a}rio do Bacanga, S\~{a}o Lu\'{i}s (MA), 65085-580, Brazil}

\begin{abstract}

This chapter of the proceedings for the {\em Ninth Meeting on CPT and Lorentz Symmetry} is dedicated to the Hamiltonian formulation of the minimal gravitational Standard-Model Extension. Some theoretical questions associated with the latter shall be reviewed. First, we recall the properties of the Hamiltonian, which was computed elsewhere, and discuss how it is linked to the modified Einstein equations. Second, we describe how the covariant and Hamiltonian formulations are shown to be consistent with each other.

\end{abstract}

\bodymatter

\section{Introduction}

Perihelion precession of Mercury, light deflection by massive bodies, and gravitational redshift strongly support General Relativity (GR) as the correct theory of gravity. However, having the viewpoint that gravity enters a quantum regime at the Planck scale, gravitational phenomena are supposed to exhibit deviations from GR predictions. These are expected to become dominant at length scales near the Planck length and can be searched for in experiments that are sensitive enough to probe them. To perform broad searches for such effects and to be able to compare measurements from different experiments with each other, the gravitational SME\cite{Kostelecky:2003fs,Bailey:2006fd} can serve as a comprehensive effective-field-theory framework, which parameterizes physics beyond GR in terms of background fields. In what follows, we would like to delve into several theoretical questions that emerge in the minimal gravitational SME.

\section{Diffeomorphism violation in the SME}

The action of the minimal gravitational SME\cite{Kostelecky:2003fs,Bailey:2006fd} is a perturbation of the Einstein--Hilbert action. We study a subset of the latter given by:
\begin{equation}
\label{eq:gravitational-sme-minimal}
S=\int_{\mathcal{M}}\mathrm{d}^4x\,\frac{\sqrt{-g}}{2\kappa}\left[(1-u){}^{(4)}R+s^{\mu\nu}{}^{(4)}R^T_{\mu\nu}\right],
\end{equation}
with $\kappa=8\pi G_N$, the determinant $g$ of the spacetime metric $g_{\mu\nu}$, the traceless Ricci tensor ${}^{(4)}R^T_{\mu\nu}$, and the Ricci scalar ${}^{(4)}R$ on the spacetime manifold $\mathcal{M}$. Furthermore, $u$ is a scalar and $s^{\mu\nu}$ is a tensor-valued background field living in the tangent bundle of $\mathcal{M}$. The tangent frames are generic and do not necessarily correspond to freely falling inertial frames. Thus, $s^{\mu\nu}$ carries spacetime indices as opposed to local Lorentz indices, whereupon a breakdown of local Lorentz invariance does not occur in the first place.

As the notion of a constant tensor-valued field on a curved manifold is obscure, the background fields must be taken as dependent on the spacetime coordinates. The behavior of these backgrounds under general coordinate transformations generated by a vector field $\xi^{\mu}$ is governed by the Lie derivatives of the backgrounds with respect to $\xi^{\mu}$, e.g., $u\mapsto u'=u+\mathcal{L}_{\xi}u$. In contrast, the backgrounds are nondynamical fields, i.e., they are neither described by an action nor do they obey any dynamical principle, but they are predetermined objects that are manually incorporated into the theory. This makes the backgrounds transform trivially under diffeomorphisms, which are the active counterparts of general coordinate transformations. Then, a diffeomorphism maps each background field onto itself, e.g., $u\mapsto u$. As a consequence, the action stated in Eq.~\eqref{eq:gravitational-sme-minimal} breaks diffeomorphism invariance explicitly. Note that there are relationships between Eq.~\eqref{eq:gravitational-sme-minimal} and Brans--Dicke theory as well as de Rham--Gabadadze--Tolley massive gravity.

The action provided by Eq.~\eqref{eq:gravitational-sme-minimal} is the foundation of the covariant formulation of the minimal gravitational SME. It gives rise to modified Einstein equations\cite{Bailey:2006fd} that involve the background fields proper and second-order covariant derivatives of the latter. The spacetime manifold is a four-dimensional entity and an understanding of the dynamical properties of this modified-gravity theory can only be achieved by investigating the modified Einstein equations, which is an immense challenge. There is a powerful formalism called the $(3+1)$ decomposition\cite{Arnowitt:1962hi} of spacetime, which is also denoted as the ADM decomposition and renders the dynamical content of such a theory much more transparent than does the covariant formulation. This formalism has turned out to be of great benefit in SME gravity.\cite{ONeal-Ault:2020ebv,Reyes:2021cpx}

The ADM decomposition foliates spacetime into a family of purely spacelike hypersurfaces $\Sigma_t$ of constant time. It is parameterized by a scalar lapse function $N$ and a shift vector with components $N^i$. The lapse function characterizes the time coordinate employed, e.g., $N=1$ for proper time. The shift vector describes a dislocation of hypersurfaces with respect to each other. Both quantities are nondynamical and are auxiliary ones, i.e., they do not contain physical information and are considered gauge degrees of freedom. This formulation brings along with itself a possible decomposition of the spacetime metric $g_{\mu\nu}$, of curvature-related quantities as well as of background fields into contributions that live in $\Sigma_t$, ones that are defined in the one-dimensional subspace orthogonal to $\Sigma_t$, and mixed components. It is then found that the physical components of the spacetime metric are actually the purely spacelike ones, $q_{ij}:=g_{ij}$, whereas $g_{0i}$ and $g_{00}$ involve $N$ and $N^i$, i.e., they cannot be physical. By applying this technique to Eq.~\eqref{eq:gravitational-sme-minimal}, the latter decomposes into three parts according to the description provided above.

The next step is to associate canonical momenta $\{\pi_N,\pi_i,\pi^{ij}\}$ with the ADM variables $\{N,N^i,q_{ij}\}$. As the physical and gauge degrees of freedom are now clearly separated from each other, a Legendre transformation applied to the ADM-decomposed Lagrangian gives rise to a Hamiltonian\cite{Reyes:2021cpx} associated with the modified-gravity theory considered in Eq.~\eqref{eq:gravitational-sme-minimal}. The latter contains a scalar piece multiplied by $N$, which is called the Hamiltonian constraint in GR. Furthermore, there is a vector-valued part contracted with $N^i$, which is often denoted as the momentum constraint. In GR, the names are chosen wisely, as these expressions are, in fact, constraints, i.e., they do not contain dynamical information. Whether or not this holds for our modified Hamiltonian\cite{Reyes:2021cpx} is still subject to ongoing investigations. Also, the situation is more involved than in GR, as the Hamiltonian depends on additional terms that do not belong to either of these two classes mentioned.

When it comes to GR, the Einstein equations ${}^{(4)}G^{\mu\nu}=\kappa T^{\mu\nu}$ with the Einstein tensor ${}^{(4)}G^{\mu\nu}$ on $\mathcal{M}$ and the energy--momentum tensor $T^{\mu\nu}$ are known to decompose into three sectors, too. For the convenient choice $N^i=0$, the purely timelike sector ${}^{(4)}G^{00}=\kappa T^{00}$ is related to the aforementioned Hamiltonian constraint and the mixed part ${}^{(4)}G^{0i}=\kappa T^{0i}$ to the momentum constraint. It is only the purely spacelike piece ${}^{(4)}G^{ij}=\kappa T^{ij}$, made up of six independent equations, that contains the physical degrees of freedom. The latter consist of four auxiliary and two propagating ones.

The Hamiltonian formulation is governed by two sets of equations:
\begin{equation}
\label{eq:hamilton-equations}
\dot{q}_{ij}=\{q_{ij},H\}\,,\quad \dot{\pi}^{ij}=\{\pi^{ij},H\},
\end{equation}
with suitably defined Poisson brackets $\{\bullet,\diamond\}$. The latter are first order in time, whereas the Einstein equations are second order in time. The first set of Hamilton's equations simply corresponds to a geometrical identity, which means that it does not contain dynamical information. In fact, it is the second set of Hamilton's equations, which is known to be equivalent to the purely spacelike part of the Einstein equations. So the covariant formalism is equivalent to the Hamiltonian approach, indeed. Both describe the dynamics and constraint structure of GR consistently with each other.

Now, our interest was to show\cite{Reyes:2022mvm} that the modified-gravity theory based on Eq.~\eqref{eq:gravitational-sme-minimal} shares the properties with GR that were summarized previously. To do so, we had to consult the modified Einstein equations for the $u$ and $s^{\mu\nu}$ sector, which had already been obtained earlier.\cite{Bailey:2006fd} The first step was to project the modified Einstein equations into $\Sigma_t$. We did so for $u$ as well as the purely timelike and spacelike sectors of $s^{\mu\nu}$. The mixed sector of $s^{\mu\nu}$ involves 3 controlling coefficients and its Hamiltonian can be brought into the form of the GR Hamiltonian by a redefinition of the canonical momentum. Thus, the mixed sector of $s^{\mu\nu}$ was interpreted as unphysical and it was discarded. After deriving the modified Einstein equations projected into $\Sigma_t$, the second step was to evaluate Hamilton's equations (\ref{eq:hamilton-equations}).

The first set was shown to be satisfied automatically for each of the three sectors. This outcome was expected, as we still worked in a pseudo-Riemannian setting. To analyze the second set of Hamilton's equations for the three sectors, Poisson brackets of the modified canonical momenta and the corresponding ADM Hamiltonians had to be computed. From a technical perspective, the computations turned out to be tedious, which is why we relied on powerful software tools. Ultimately, we were able to demonstrate the equivalence between the covariant and Hamiltonian approaches to Eq.~\eqref{eq:gravitational-sme-minimal}. This finding corroborates the use of nondynamical background fields as a parameterization of explicit diffeomorphism violation. If there were internal inconsistencies in such a theory, they would presumably have been revealed in this analysis.

\section*{Acknowledgments}

The author is indebted to FAPEMA Universal 00830/19, CNPq Produtividade 312201/2018-4, and CAPES/Finance Code 001.

\end{document}